\documentclass[journal=jacsat,manuscript=article]{achemso}
\usepackage[version=3]{mhchem} 

\author{Bj\"{o}rn C. P. Sturmberg}
\email{b.sturmberg@physics.usyd.edu.au}
\affiliation[CUDOS and IPOS, School of Physics, University of Sydney, 2006, Australia]
{University of Sydney}
\author{Kokou B. Dossou}
\author{Lindsay C. Botten}
\author{Ara A. Asatryan}
\author{Christopher G. Poulton}
\affiliation[CUDOS, School of Mathematical Sciences, University of Technology Sydney, Sydney, 2007, Australia]
{University of Technology Sydney}
\author{Ross C. McPhedran}
\author{C. Martijn de Sterke}
\affiliation[CUDOS and IPOS, School of Physics, University of Sydney, 2006, Australia]
{University of Sydney}

\title{Optimizing Photovoltaic Charge Generation of Nanowire Arrays: A Simple Semi-Analytic Approach}

\begin{document}
\begin{abstract}
Nanowire arrays exhibit efficient light coupling and strong light trapping, making them well suited to solar cell applications.
The processes that contribute to their absorption are interrelated and highly dispersive, so the only current method of optimizing the absorption is by intensive numerical calculations.
We present an efficient alternative which depends solely on the wavelength-dependent refractive indices of the constituent materials.
We choose each array parameter such that the number of modes propagating away from the absorber is minimized while the number of resonant modes within the absorber is maximized.
From this we develop a semi-analytic method that quantitatively identifies the small range of parameters where arrays achieve maximum short circuit currents. This provides a fast route to optimizing NW array cell efficiencies by greatly reducing the geometries to study with full device models.
Our approach is general and applies to a variety of materials and to a large range of array thicknesses.
\end{abstract}

\textbf{Keywords:} photovoltaics, solar energy, nanowire arrays, light trapping, nanophotonics, diffraction gratings, effective medium approximation

A major challenge in solar energy research is to reduce the cost-per-Watt of photovoltaic cells. This challenge may be met by increasing the efficiency, as in the case of multi-junction cells \cite{King2009}, or by reducing costs through the use of less expensive materials \cite{Todorov2010,Liu2013}. The cost-per-Watt can also be reduced by using established materials in much thinner films than standard cells \cite{Catchpole2006b,Tsakalakos2008}. This last approach greatly reduces costs, but also reduces the efficiency if the cell design is not modified. The efficiency decrease is largely due to optical losses; not only from increased transmission due to incomplete absorption, but also from increased reflection at the front surface, which can no longer be coated with standard multiple micrometer-thick anti-reflective patterns. Structuring the thin absorbing layer into vertically aligned Nanowire (NW) arrays has been shown to address both these losses simultaneously, by increasing the {\it light coupling} into the cell and enhancing the absorption through {\it light trapping}. Further advantages of NWs include compatibility with cheap flexible substrates \cite{Fan2009} and radial p-n junctions \cite{Kayes2008,Garnett2008,Mariani2013a}. NWs also reduce the lattice matching constraints that limit material combinations in multi-junction devices \cite{Endo2013}. These structures have therefore attracted intense theoretical \cite{Sturmberg2011,Lin2009,Alaeian2012,Wang2012b} and experimental \cite{Gunawan2010,Hu2007,Fan2009,Heurlin2011,Huang2012,Madaria2012,Howell2013} research with efficiencies of 13.8\% being reported \cite{Wallentin2013a}.

Optimizing the photovoltaic performance of NW arrays is challenging because their optical and electronic properties depend strongly upon the arrays' geometry. 
Typically, simulations are first carried out to assess the optical charge carrier generation and then the charge carrier collection performance is calculated separately. Recently, techniques for combined opto-electronic modeling of nanostructured photovoltaics have been developed that recursively calculate the generation, collection and recombination of charge carriers \cite{Bermel2014}.
All of these numerical approaches are limited to grid searches of small parts of parameter-space, wherein they locate local optima.
Finding a general optimum, or fully understanding the behaviour of NW arrays using such means, is however impracticable due to the range of parameters.
In particular, the high sensitivity of the charge carrier profile to experimental factors such as material quality, fabrication techniques, passivation method and device design prohibits the carrier collection efficiency from being incorporated into a general optimization routine. 
These issues also limit the predictive information that can be gained from experiments. For example, the fill factors of current state of the art III-V NW array solar cell vary between 25-77\% \cite{LaPierre2013b}.
It is therefore valuable to fully understand the charge generation aspects of NW arrays before modeling the charge collection of arrays. A compromise between maximal charge generation and efficient charge collection can then be found to maximize the photovoltaic energy conversion efficiency.

The charge generation of NW arrays is directly related to their optical absorption \cite{Kailuweit2011}, which is less dependent on material processing.
The absorption in turn however is driven by numerous competing optical effects, including the excitation of guided resonance modes and higher diffraction orders \cite{Sturmberg2011,Lin2009,Alaeian2012,Wang2012b}. Since these are resonant effects, they are highly dispersive across the broad bandwidth of the solar spectrum and it is unclear how to best balance their effects to maximize the total absorption. This is made more difficult by each wavelength contributing an amount given by the solar spectrum.
These factors all combine to produce a challenging optimization problem that is currently addressed by fully numerical studies of specific sections of parameter-space \cite{Lin2009,Alaeian2012,Hu2007,Huang2012}.

\subsection{Semi-Analytic Approach}

\begin{figure}
\begin{center}
   \includegraphics[width=0.9\linewidth]{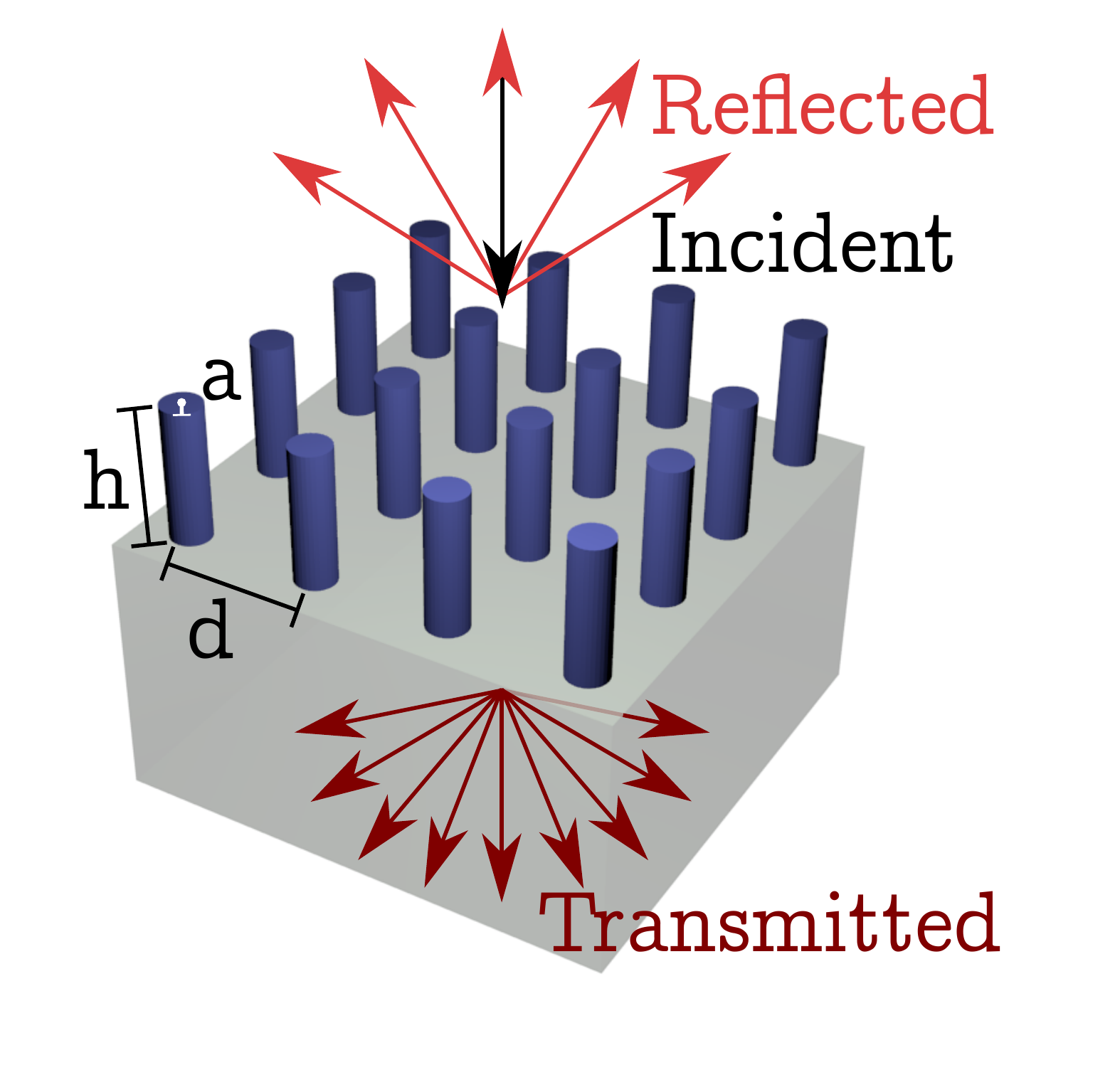}
\end{center}
\vspace{-7mm}
\caption{Geometry of vertically aligned nanowire arrays showing light incident normally and diffracted into the zeroth and higher orders in reflection and transmission. While our optimization method is independent of lattice type numeric simulations use a square lattice as shown.}
\label{Opt_geo}
\end{figure}

Here we present an approach that allows the optimal design parameters for maximal charge generation to be evaluated semi-analytically in negligible time for a given material and thickness.
This provides an ideal starting point for numerical and experimental studies of the electrical performance of NW geometries.
We do this by constructing an integrated theoretical framework that relates the reflection, transmission and absorption of NW arrays to their geometric parameters (\ref{Opt_geo}): radius $a$, period $d$, thickness $h$ and volume fraction $f$. 
In our numerical calculations we consider square lattices for which $f~=~\pi~a^2/d^2$, however studies have reported a weak influence of lattice type on the absorption of NW arrays \cite{Alaeian2012,Li2012} and our analytic model is independent of the lattice type.
From these relationships we develop a simple method that quantitatively identifies the small region of optimal parameter-space for charge generation.

\begin{figure}
\begin{center}
   \includegraphics[width=0.95\linewidth]{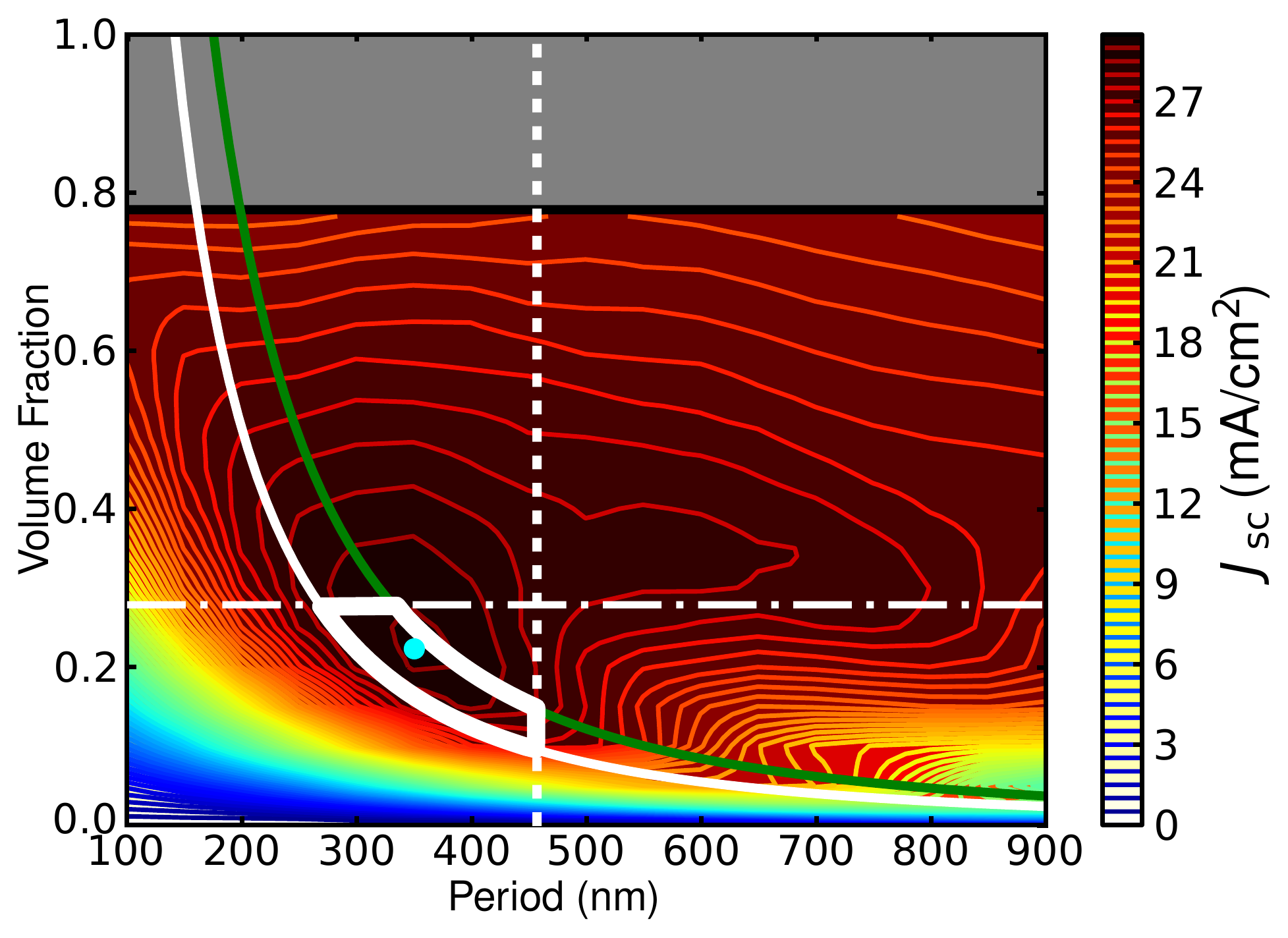}
\end{center}
\vspace{-7mm}
\caption{Numerically calculated short circuit current of InP NW arrays across the parameter-space $100~\text{nm}~<~d~<~900~\text{nm}$, $0.005~<~f~<~0.780$. The optimal array is marked with a blue dot and has $J_{\rm sc}~=~29.2$~mA/cm$^2$. Superimposed are the results of our semi-analytic optimization with the predicted optimal region emphasised with thick curves.}
\label{InP}
\end{figure}

For each choice of material system and thickness we analyse the parameter-space presented in \ref{InP}. Here the period and volume fraction are indicated on the axes, and curves of constant radius are of the form $f~\propto~1/d^2$. Arrays whose NWs intersect are excluded by the grey region, which for square arrays is $f~>~\pi/4$. The colored contours of \ref{InP} indicate numerically calculated short circuit currents $J_{\rm sc}$ of NW arrays where the array that produce maximal $J_{\rm sc}$ is marked with a blue dot.
Overlaid on the numerical results, with white and green curves, are the predictions of our semi-analytic optimization.
This optimization method is briefly described as dividing parameter-space into regions that have either too high a reflectance (above the horizontal dot-dashed line), too much transmission (right of the vertical dashed line), or do not support enough resonant absorption modes (outside of the region between the white and green curves). The optimal NW design must therefore lie within the region highlighted in thick white curves.
The considerations dictating these reductions make use of the coupled wave argument developed by Yu et al. \cite{Yu2011}. That is, that the absorption is highest when the number of outward propagating modes is minimized, while the number of resonant modes within the absorbing layer is simultaneously maximized.
Here we extend this approach by considering the modal amplitudes, and by integrating across the solar spectrum to calculate parameter constraints in terms of the maximum short circuit current.

We illustrate our method using InP NWs surrounded by an air background ($n~=~1$), placed upon a semi-infinite SiO$_2$ substrate under an air superstrate.
We stress however that the method applies to a wide variety of materials including direct and indirect bandgap semiconductors. Results for GaAs, silicon and germanium are contained in the Supporting Information. We first examine the effect of radius, period and volume fraction while fixing $h~=~2.33~{\rm \mu m}$, and thereafter show that the thickness determines the relative importance of reflection {\sl vs} transmission losses.
Our semi-analytic predictions show excellent agreement for the optimal array parameters when compared to full simulations.

\textbf{Effect of NW Radius on Absorption Resonances.}
We begin by studying the role of the NW radius.
It has been shown that NW arrays support guided resonance modes, each of which drive strong absorption peaks.
These modes have been described in the literature as either leaky optical fibre modes \cite{Anttu2010,Wang2012b} or as \textit{Key (Bloch) Modes} (KMs) \cite{Sturmberg2011}.
These descriptions are equivalent for sparse arrays where the NW surfaces are far apart, but differ for dense arrays where the effect of the lattice, which is included only in the Bloch mode formulation, becomes significant (see Supporting Information for derivation of equivalence in the sparse limit). Though here we consider perfectly cylindrical NWs, the lower order modes of NWs depend on the cross-sectional area so that minor perturbations in shape do not alter the dipolar nature of the fields, and for our purposes here the same equations can be applied. A fundamental property of waveguides is that the number of bound modes increases with the dimension of the cross-section. Applied to NWs this means that the number of KMs grows with increasing radius.
This suggests a clear optimization goal; choose the NW radius as large as possible, so as to maximize the number of absorption resonances.

\begin{figure}
\begin{center}
   \includegraphics[width=0.8\linewidth]{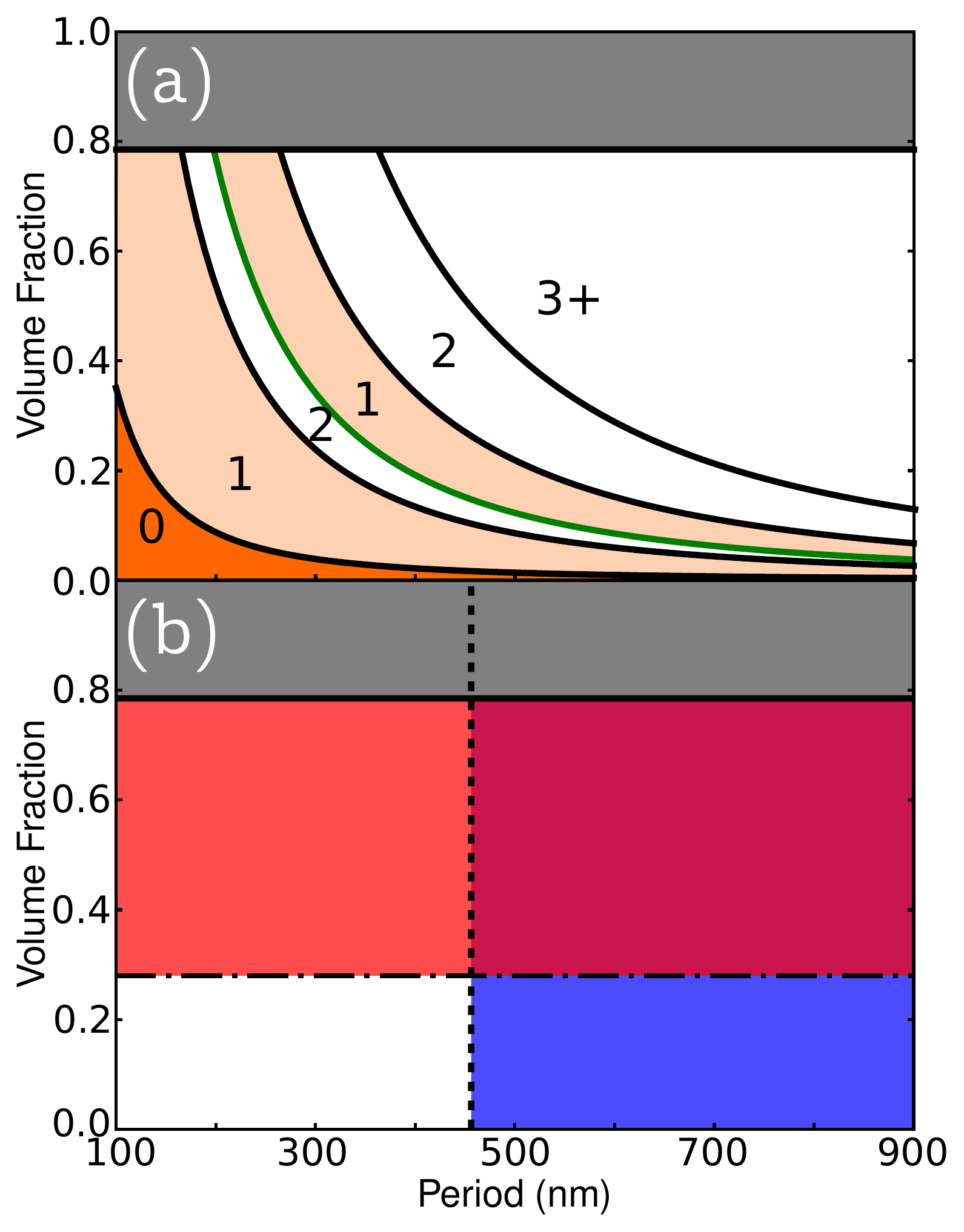}
\end{center}
\vspace{-7mm}
\caption{(a) NW array parameter-space showing the NW radii at which KMs enter (black curves) and leave (red curve) the absorption spectrum. The number of KMs within the absorption spectrum is shown for each region, where darker shades of orange indicate fewer KMs.
(b) Constraints placed on the optimal region of parameter-space due to high transmission (blue area) and high reflection (red area). The region of high transmission and reflection is coloured magenta.
Optimal NW array geometries lie in the union of the white regions of (a) and (b).}
\label{predict_InP}
\end{figure}

However, absorption resonances only contribute to the absorption of the solar cell if they occur at wavelengths where the material absorbs, and where there is incoming solar radiation. This is the range $\lambda_l~<~\lambda~<\lambda_g$, where $\lambda_l~=~310$~nm is the lower limit of the solar spectrum and $\lambda_g$ is the bandgap wavelength of the absorbing material, which for InP is $\lambda_g = 922.5$~nm.
It is known that the symmetry of the incident light allows only HE$_{1m}$ fiber modes to be effectively excited \cite{Wang2012b,Anttu2013c}. We have found that the strongest excitations occur at frequencies slightly above the $\text{Re}(\beta_{\rm z})~=~0$ cut-off of these modes, where $\beta_{\rm z}$ is the propagation constant of the mode. Hence, the strongest absorption occurs when 
\begin{equation}
\label{cut-off}
\dfrac{\epsilon_1 J^{'}_1(k_1a)}{k_1 J_1(k_1a)} - \dfrac{\epsilon_2 H^{(1)'}_1(k_2a)}{k_2 H_1^{(1)}(k_2a)} = 0.
\end{equation}
which can be derived from the standard transcendental equation for HE modes in the limit $\beta~\to~0$. Here $\epsilon_i$ and $k_i~=~\sqrt{\epsilon_i\omega^2/c^2~-~\beta_{\rm z}^2}$ are the permittivities and transverse wavenumbers in the NWs ($i~=1$) and background ($i~=2$) respectively, $J_1$ is the first order Bessel function and $H^{(1)}_1$ is the first order Hankel function of the first kind. Incidentally, this expression is identical to the one obtained for the Key Modes, the Bloch modes which are known to dominate the absorption in NW arrays, in the limit in which the lattice period becomes arbitrarily large ({\sl i.e.}, as $S_0~\to~0$ in Eq.~(10) in Sturmberg {\sl et al.}\cite{Sturmberg2011}). The fundamental HE$_{11}$ fiber mode does not have a cut-off. It therefore exists for all radii at all wavelengths, but does not contribute a large absorption peak and is not counted in \ref{predict_InP}(a).
For the derivation of Eq.~1 and the relationship between KMs and HE$_{1m}$ fiber modes see the Supporting Information.

Using Eq.~(1) we scan through possible NW radii recording when Key Modes enter and leave the spectrum: $\text{Re}(\beta_{\rm z}(\lambda_l))~=~0$, $\text{Re}(\beta_{\rm z}(\lambda_g))~=~0$ respectively.
These radii are represented in \ref{predict_InP}(a) as black curves where an additional mode enters the spectrum and as green curves where a mode leaves the spectrum. Between the curves the number of KMs is constant, as labelled in \ref{predict_InP}(a). From this analysis we require the optimal arrays to be in one of the regions with at least 2 KMs, such as immediately to the left of the green curve or towards the top right of the figure. There is a preference for small radii because the fields of higher order modes, which are supported by larger NWs, have a greater number of nodes within the NW and therefore couple less well to the incident plane waves.
For diagrammatic purposes we include only the first 3 KMs in \ref{predict_InP}(a).

Having identified the light trapping advantages of larger NWs we now consider two processes that impose upper limits to the optimal NW radius. These arise because the NWs are arranged in a periodic lattice; thus increasing the radius we can either keep the volume fraction constant and adjust the period, or keep the period constant and adjust the volume fraction. These correspond, respectively, to moving horizontally and vertically through \ref{InP}. In each case we observe that the absorption initially grows with increasing radius, but eventually decreases. It is by uncovering the origins of this behavior that we can define the vertical and horizontal restrictions shown in \ref{InP}.

\begin{figure}
\begin{center}
   \includegraphics[width=0.95\linewidth]{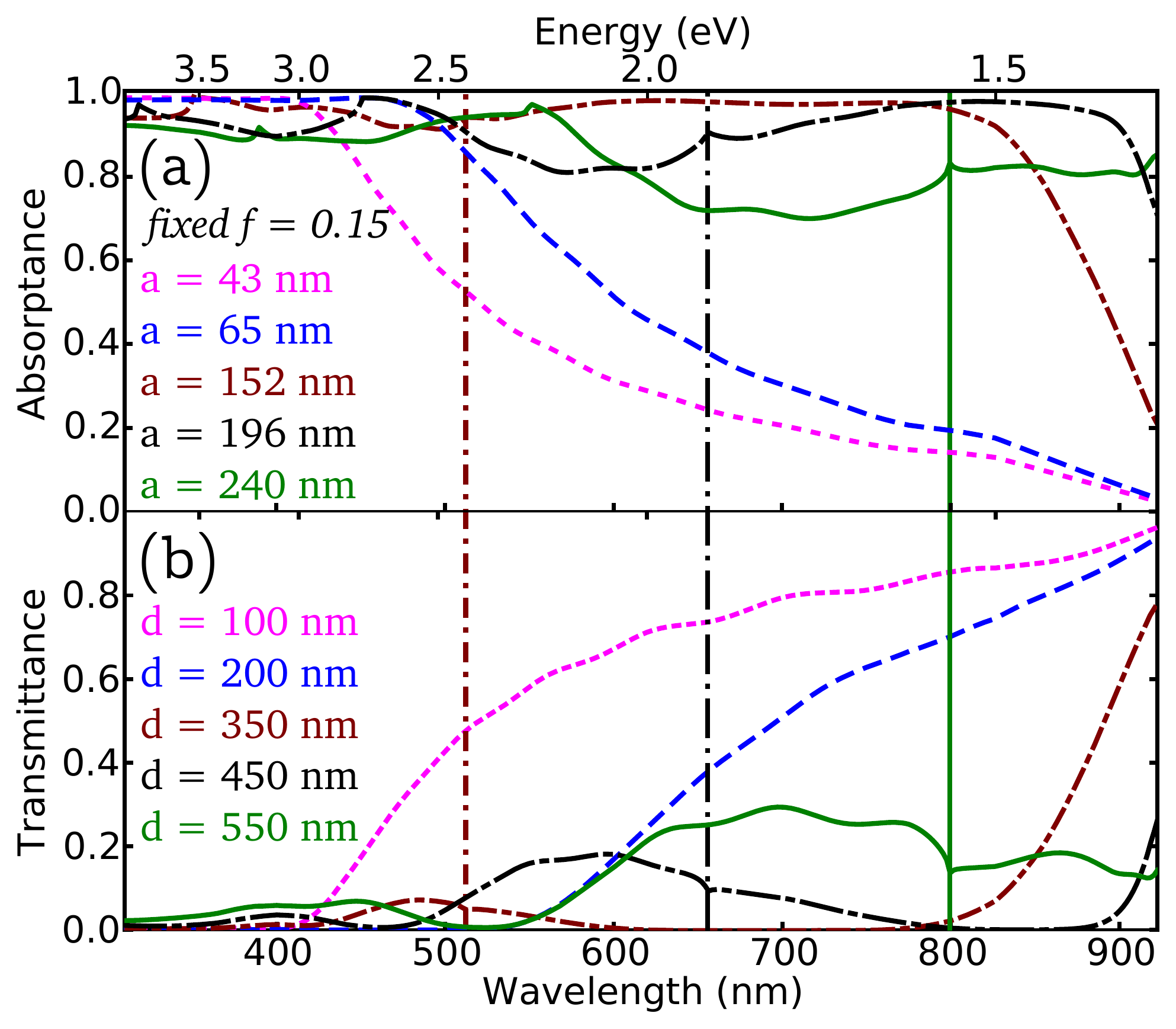}
\end{center}
\vspace{-7mm}
\caption{Absorption (a) and transmission (b) spectra as NW radius is increased with a fixed volume fraction $f~=~0.15$. Vertical lines mark the shortest wavelength Wood anomalies in the substrate $\lambda_{WA}~=~n_{\rm sub}d$.}
\label{fix_f}
\end{figure}

\textbf{Effect of Array Period on Transmission Channels.}
We investigate the role of increased period in \ref{fix_f}, where we show the absorption and transmission spectra of InP arrays with radii in the range $43 - 240$~nm and fixed volume fraction $f~=~0.15$, \textit{i.e.} moving horizontally in \ref{InP}.
In \ref{fix_f}(a) the absorption initially increases with radius, but once $a~>~152$~nm ($d~>~350$~nm) the absorption of long-wavelengths is significantly reduced.
It is clear from \ref{fix_f}(b) that this decrease is due to a rise in transmission, while the reflectance is relatively constant for all arrays, with $R~<~7\%$ across the spectrum (see Supporting Information).
The increase in transmission is caused by the periodically varying refractive index of the NW array exciting non-zero diffraction orders in the substrate, which increases the number of channels propagating energy away from the array from $1$ to $5$ for wavelengths below the zero order Wood anomaly. This anomaly occurs at $\lambda~=~\lambda_{WA}$, which is indicated in \ref{fix_f} by vertical lines in the line style of the corresponding spectrum. In transmission the shortest wavelength Wood anomaly occurs at $\lambda_{WA}~=~n_{\rm sub}d$, where $n_{\rm sub}$ is the refractive index of the substrate.
For $\lambda~>~\lambda_{WA}$ all non-zero diffraction orders are evanescent, and as such do not contribute to the energy flow.
The excitation of higher diffraction orders therefore adds a qualifier to the optimization; one should incorporate large NW radii, but do so while imposing an upper limit on the period.

The strictest limit to place on the period is that no higher diffraction orders are allowed to exist within the solar spectrum, which limits the period to $d~<~310~{\rm nm}/n_{\rm sub}~=~213$~nm for a SiO$_2$ substrate.
This criterion however fails to consider the strong absorption of short wavelengths that do not reach the substrate, which renders the number of transmission channels irrelevant for these wavelengths. We must therefore choose a longer wavelength $\lambda_{\rm av}$ from which we can derive the maximum allowable period, $d_{\rm max}~\equiv~\lambda_{\rm av}/n_{\rm sub}$.

Our approach is to choose $\lambda_{\rm av}$ to be an averaged wavelength, on the short wavelength side of which the non-zero transmission channels can be tolerated, while there is only single transmission channel at longer wavelengths.
To account for the highly non-uniform spectral intensity of the solar spectrum we weight each above-bandgap wavelength by the number of solar photons incident at that wavelength $\xi(\lambda) = I(\lambda)\lambda/hc$, where $I(\lambda)$ is the irradiance as given in the ASTM Air Mass 1.5 spectrum \cite{ASTM}, $h$ is Planck's constant and $c$ is the speed of light in vacuum:

\begin{equation}
\label{average}
\lambda_{\rm av} =  \dfrac{\int_{\lambda_l}^{\lambda_g} \xi(\lambda) \lambda d\lambda} {\int_{\lambda_l}^{\lambda_g} \xi(\lambda) d\lambda}.
\end{equation}

For InP this gives $\lambda_{\rm av}~=~665$~nm. We take an analogous weighted average of the substrate refractive index, replacing $\lambda$ in the numerator with $n(\lambda)$, which gives $n_{\rm sub}~=~1.46$. 
The maximum allowed period is therefore $d_{\rm max}~=~456~\text{nm}$, which is represented in \ref{predict_InP}(b) by the start of the blue region.

\begin{figure}
\begin{center}
   \includegraphics[width=0.95\linewidth]{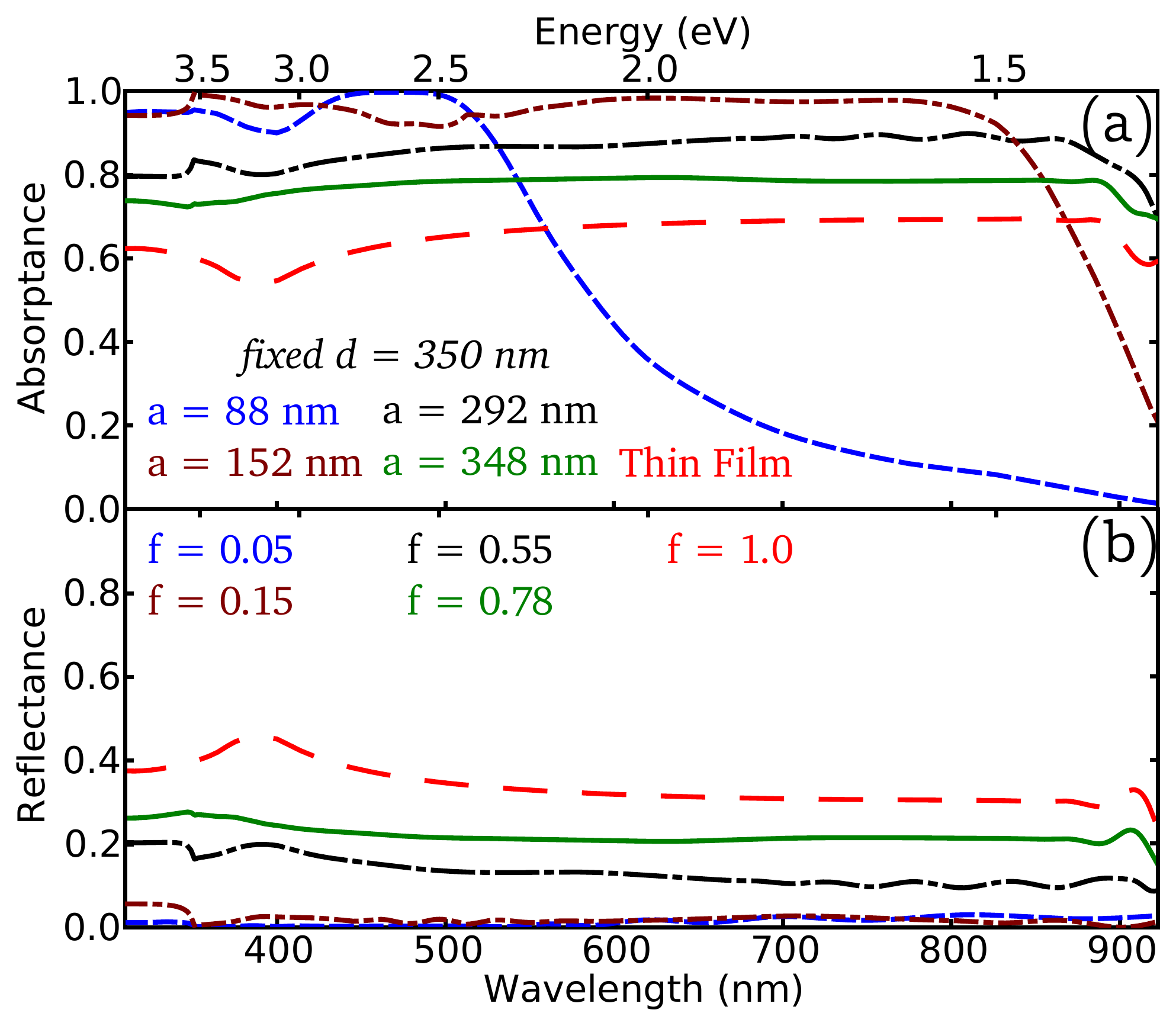}
\end{center}
\vspace{-7mm}
\caption{Absorption (a) and reflection (b) spectra as NW radius is increased with a period $d~=~350$~nm. The spectra of a homogeneous film of equal thickness is shown in red ($f~=~1.0$).}
\label{fix_d}
\end{figure}

\textbf{Effect of Volume Fraction on Reflectance.}
Having examined the effect of increasing NW radii within lattices of increasing periods ({\sl i.e.} moving horizontally in \ref{InP}), we now move vertically through \ref{InP}.
\ref{fix_d} shows the absorption and reflection spectra for arrays with fixed period $d~=~350$~nm and radii increasing from $a~=~88$~nm to $a~=~348$~nm ($0.05~\le~f~\le~0.78$).
The spectra of a homogeneous film of equal thickness ($f~=~1.0$) is also shown to emphasise the excellent anti-reflective light coupling performance of the NW arrays.
The transmittance of these arrays, shown in the Supporting Information, is relatively constant once $f~\ge~0.55$.
In contrast to \ref{fix_f}(a), the absorption in \ref{fix_d}(a) decreases uniformly across the spectrum once $a~>~152$~nm and does so due to a uniform increase in reflectance.
The Wood anomaly in reflection that occurs at $\lambda_{\rm WA}~=~n_{\rm Air}d = 350~{\rm nm}$ produces only a minor enhancement in short wavelength reflection, which is unaffected by the increase in $f$ (since $d$ is constant). The increased reflectance is therefore due not to a larger number of reflected propagating orders, but to a larger amplitude of the zeroth reflection order. This trend is not surprising: as the InP volume fraction increases, so does the average refractive index, and therefore so does the Fresnel reflection.

To incorporate the influence of Fresnel reflections on the optimization we determine an upper bound on the volume fraction $f_{\rm max}$ that does not impede the integrated absorption. This is a subtle procedure that depends on the thickness of the absorbing layer: thick layers can be expected to absorb most of the light which enters, and minimizing reflection is therefore paramount. In contrast, thin layers with limited absorption can tolerate more reflection if it leads to higher absorption. Rather than dealing with the absorption, we use the transmission as a proxy since, for example, thick layers with strong absorption have little transmission. We therefore define the largest volume fraction allowed as that for which the average reflection equals the average transmission.
The reflectance $R(\lambda)$ and transmittance $T(\lambda)$ are estimated using the Fresnel equations with the NW array replaced by a thin film of equal thickness composed of an effective medium with permittivity $\epsilon_{\rm eff}(f,\lambda)$.
The critical volume fraction $f_{\rm max}$ is then defined as the $f$ such that $R_{\rm av}(\epsilon_{\rm eff}(f))~=~T_{\rm av}(\epsilon_{\rm eff}(f))$, where the average reflectance and transmittance is calculated as in Eq.~\ref{Rave} (see Supporting Information for derivation).
For InP NW arrays with $h~=~2.33~\mu$m, $f_{\rm max}~=~0.28$, which is indicated in \ref{predict_InP}(b) by the red shading of $f~>~0.28$.

\begin{equation}
\label{Rave}
R_{\rm av}(f) =  \dfrac{\int_{\lambda_l}^{\lambda_g}\xi (\lambda) R(f, \lambda) d\lambda} {\int_{\lambda_l}^{\lambda_g} \xi(\lambda) d\lambda}.
\end{equation}

It is important to note that this limit relies on the calculation of effective indices for structures that include significant loss. Existing effective index models, such as Maxwell-Garnett, poorly approximate the complex effective indices for these structures. It has recently been proposed in studies on mesoporous thin films, that the real and imaginary components of the effective indices of lossy structures may be best calculated independently from different effective index formulations. As in these studies \cite{Hutchinson2010,Navid2008}, we found such an approach to best replicate the observed reflectance \textit{and} transmission spectra, and subsequently to most reliably predict $f_{\rm max}$. We calculate $\text{Re}(\epsilon_{\rm eff})$ using the Bruggeman formulation \cite{Bruggeman1935} and $\text{Im}(\epsilon_{\rm eff})$ using the Volume Averaging Theory \cite{DelRio2000a,DelRio2000}.

\subsection{Comparison to Numeric Simulations}

To make a final prediction of the optimal array parameters we combine the three arguments developed above.
This corresponds to placing the restrictions on period and volume fraction of \ref{predict_InP}(b) onto the radius goals of \ref{predict_InP}(a).
In doing so we remove the top right region of parameter-space that supports three KMs, as well as the slightly smaller radius region that supports two KMs. The optimal regions is therefore predicted to be just to the left of the green curve, where two KMs are supported without the excitation of too many transmission channels or too high a top surface reflectance.
To verify this prediction we calculate the short circuit current $J_{\rm sc}$ of InP NW arrays across the parameter-space of $100~\text{nm}~<~d~<~900~\text{nm}$, $0.005~<~f~<~0.780$ with $h~=~2.33~\mu$m. We sampled this parameter-space with $993$ simulations.

\ref{InP} shows these numerical results as well as the semi-analytic predictions for $f_{\rm max}$, $d_{\rm max}$ and the radius limits within which 2 KMs are supported. The agreement is excellent, with the optimal arrays located under $f_{\rm max}$ to the left of $d_{\rm max}$ and above the white solid curve ($a~=~80$~nm) that indicates the presence of the second KM resonance within the absorption range. 
The optimal array was found to have a period of $d~=~350$~nm, a volume fraction of $f~=~0.22$ and had a short circuit current of $J_{\rm sc}~=~29.2$~mA/cm$^2$.
The optimum array is marked with a blue dot and is located to the left of the green curve ($a~=~99$~nm), where the first KM leaves the absorption range.
The near optimal arrays that are in close proximity to the green curve, but on its right side, arise because these arrays still have a large fraction of the KM resonant absorption peak within the absorption range.
The contours of other regions also substantiate our approach; the horizontal contours above the $f_{\rm max}$ line indicate a volume fraction dominated (period independent) decrease in absorption, while the parameter-space with small $f$ and large $d$ exhibits near vertical contours, demonstrating the strong influence of period here.

\begin{figure}
\begin{center}
   \includegraphics[width=0.95\linewidth]{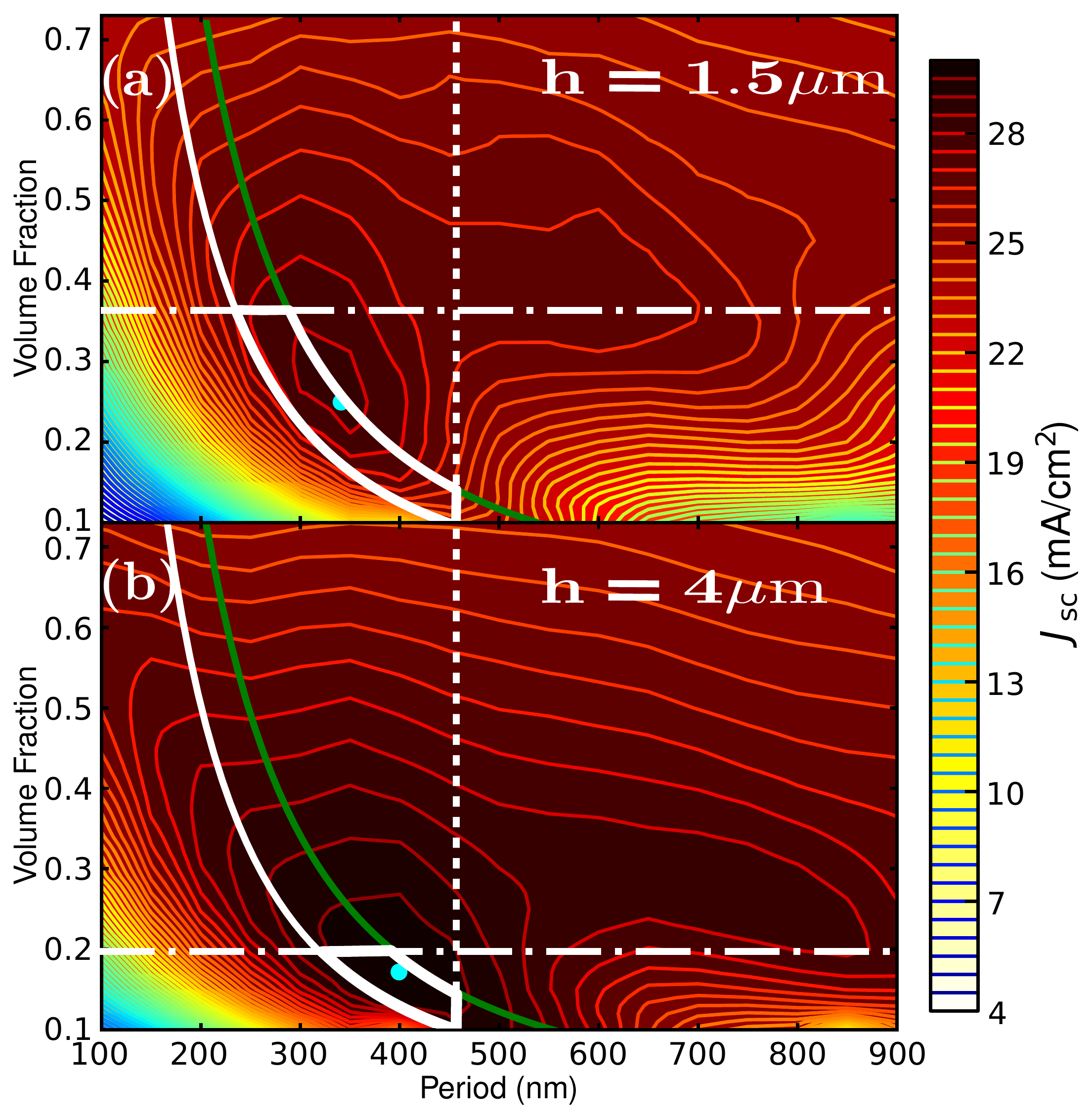}
\end{center}
\vspace{-7mm}
\caption{Numerically calculated short circuit currents $J_{\rm sc}$ of InP NW arrays of height (a) $h~=~1.5~\mu$m, (b) $h~=~4~\mu$m. The optimal arrays are marked with blue dots and have (a) $J_{\rm sc}~=~28.2$~mA/cm$^2$, (b) $J_{\rm sc}~=~29.9$~mA/cm$^2$. Superimposed are the results of our semi-analytic optimization with the predicted optimal region emphasised with thick curves.}
\label{heights}
\end{figure}

\textbf{Dependence on Array Thickness.}
The above analysis has been limited to arrays of fixed thickness, with the comparison in \ref{InP} being for $h~=~2.33~\mu$m. It is however straightforward to extend the approach to other thicknesses, requiring only $f_{\rm max}$ to be re-evaluated, due to the dependence of $T(\epsilon_{\rm eff}(f))$ on the array thickness.
In \ref{heights} we show the semi-analytic predictions for arrays of (a) $h~=~1.5~\mu$m and (b) $h~=~4~\mu$m along with the corresponding numerical calculations for $J_{\rm sc}$. The predictions continue to agree well with the numerical results, as the optimum shifts to lower $f$ with increased $h$. This is well approximated by the balancing of transmission vs. reflection through the effective medium, as demonstrated from the semi-analytic horizontal lines.

Comparing our calculated short circuit currents to the best reported experimental results, we find that for $h~=~1.5~\mu$m our maximum $J_{\rm sc}$ of $28.2$~mA/cm$^2$ is $4.2$~mA/cm$^2$ greater than that reported by Wallentin et al. \cite{Wallentin2013a} for InP NWs of equal thickness.
While some of this difference is accounted for by the absence of carrier losses in our optical simulations, the $J_{\rm sc}$ we calculate for the geometry they fabricated is only $1.3$~mA/cm$^2$ greater than their experimentally measured value. This leaves almost $3$~mA/cm$^2$ to be gained by optical optimization. The geometry of Wallentin et al. would gain this improvement by doubling the volume fraction from $0.12$ to $0.25$, while keeping the radius of the NWs at the already optimal $180$~nm.

We chose a thickness of $h~=~4~\mu$m in \ref{heights}(b) because this is the thickness of the record efficiency planar InP solar cell \cite{Keavney1990,Green2012}. This planar cell has $J_{\rm sc}~=~29.5$~mA/cm$^2$, while the optimal NW array ($d~=~400$~nm, $f~=~0.17$) has $J_{\rm sc}~=~29.9$~mA/cm$^2$.
This would suggest that, at this thickness, a planar structure will be superior to a NW array once carrier losses are included in fabricated NW arrays.
However the record planar cell contains an anti-reflective coating and a Zn-Au back reflector, neither of which are present in our NW simulations. Given the NW arrays excellent anti-reflection properties we do not introduce an anti-reflective layer, however when we include an Au back reflector the NW arrays obtain $J_{\rm sc}~=~30.5$~mA/cm$^2$ for thicknesses of $4~\mu$m. Consistent with our semi-analytic optimization the optimal arrays that included back reflectors have larger periods because there are no longer any diffraction orders excited in the substrate. The only waves that carry energy away from the NWs are in the air superstrate such that $d_{\rm max}~=~\lambda_{\rm av}/n_{\rm air}$.
These results suggest that it may be possible for NW arrays to outperform even the record planar cell that consumes $10$ times the amount of InP.

We note that the results of \ref{heights} were calculated simultaneously along with $49$ other heights across the range of 1 -- 50$~\mu$m. This was done with negligibly increased computational expense by virtue of the generalized scattering matrix method \cite{Dossou2012} as implemented in the freely available `EMUstack' simulation package \cite{EMUstack}.
For all thicknesses in this range the numerically calculated optima lie within the region predicted by our optimization method.

\subsection{Conclusion}
We have constructed a simple optimization method to maximize the photovoltaic charge generation of NW arrays. Our method depends solely on the refractive indices of the constituent materials and can be evaluated essentially instantaneously, providing a refined starting point for device modeling of NW array solar cells.
The physical premise of our approach is to maximize the light trapping of the structure while simultaneously enhancing the light coupling into the solar cell.
To achieve optimal light trapping we showed that the NW radius should be increased so as to support more resonant modes and the period should remain less than $d_{\rm max}$ to limit the number of outwardly propagating plane waves excited.
Meanwhile to enhance the light coupling into the array the volume fraction must be restricted to $f~<~f_{\rm max}$.
From this simple Ansatz we developed expressions that quantitatively define the region of parameter-space where NW arrays have optimal $J_{\rm sc}$.

\begin{figure*}
\begin{center}
   \includegraphics[width=0.95\linewidth]{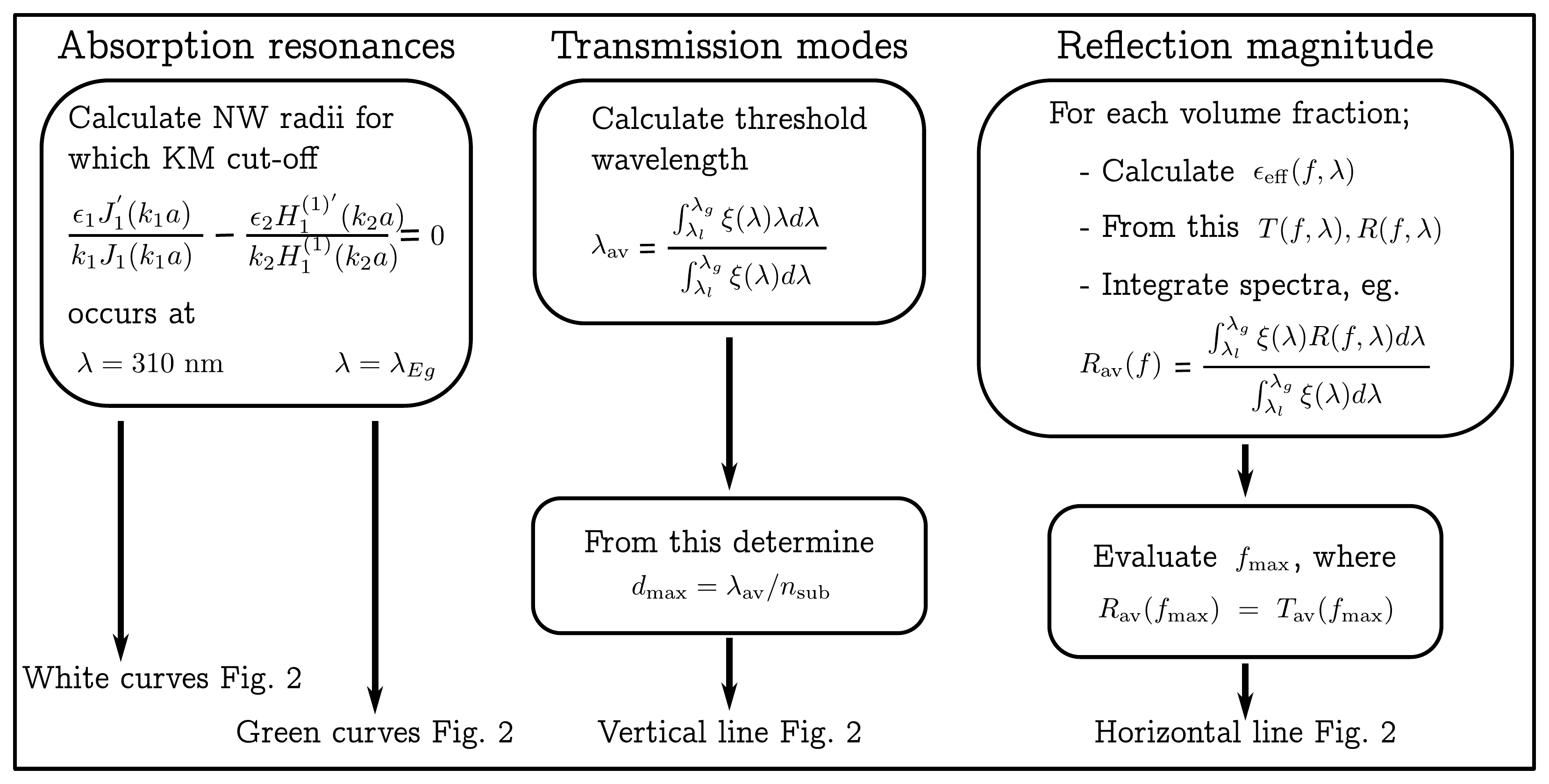}
\end{center}
\vspace{-7mm}
\caption{Flow chart outlining our semi-analytic optimization procedure. The number of resonant modes within the spectrum is found by first calculating where additional Key Modes enter the absorption range (plotted as white curves), and then calculating where they leave this range (green curves). Next, the upper bound on the period is determined and marked with a vertical line. The optimal region is then enclosed by the horizontal line which limits the acceptable volume fraction of the array above which the reflection is too great.}
\label{summary}
\end{figure*}

\ref{summary} outlines our optimization method, summarizing the equations that quantify the limits of the optimal parameter-space. By following this procedure the NW array geometries that maximize charge generation can be rapidly identified, guiding further electronic modeling and experimental investigations to simultaneously optimize the charge collection efficiency.

The construction of these arguments requires a fine balance between making simplifying approximations, and retaining the essential physics of the various optical effects.
In the calculation of $f_{\rm max}$ for example, the essential physics of balancing reflection losses against transmission losses was accentuated, while all other effects, such as the modes of the NWs, were omitted through the use of an effective index treatment.
Similarly, the derivation of $d_{\rm max}$ was based on a simple physical argument about the excitation of diffraction orders, but required the considered selection of a critical wavelength $\lambda_{\rm av}$.
Although the analysis presented here is restricted to normal incidence, it is straightforward to generalize the expressions to off-normal incidence, and we note that previous studies of NW arrays have found that the integrated absorption of NW arrays is altered little for non-normal angles of incidence up to $40^{\rm o}$ \cite{Sturmberg2011,Lin2009}.

The results contained in the Supporting Information for GaAs, silicon and germanium NW arrays show that the presented analytic optimization is robust to the very different refractive indices of these materials, including large dispersive variations in absorption coefficients. These differences produce distinct semi-analytic predictions and unique $J_{\rm sc}$ contour topologies. For example, silicon arrays of thickness $h~=~2.33~\mu$m have an optimal volume fraction of $f~=~0.7$ due to silicon's low absorptivity and an optimal period of $d~=~500$~nm due to its larger bandgap.

\acknowledgement
This work was supported by the Australian Renewable Energy Agency, and the Australian Research Council Discovery Grant and Centre of Excellence Schemes.
Computation resources were provided by the National Computational Infrastructure, Australia.
We acknowledge useful discussions with Dr T. White and Dr K. Catchpole.

\subsection{Supporting Information}
Supporting information contains comparisons of our semi-analytic optimization to numerically calculated values of $J_{\rm sc}$ for silicon, germanium and GaAs NW array solar cells. It also contains the reflection and transmission spectra complementing Figures~4,~5, the derivation of the dispersion relation of KMs and HE$_{1m}$ fiber modes which lead to Eq.~1, and the derivation of $f_{\rm max}$.
This material is available free of charge via the Internet at http://pubs.acs.org.

\bibliography{Analytic_Optimisation_Paper.bbl}

\end{document}